\documentstyle[psfig]{laa}

\def\ppap{
\mathrel{\hbox{\rlap{\hbox{\lower4pt\hbox{$\sim$}}}\hbox{$<$}}}}
\def\qqap{
\mathrel{\hbox{\rlap{\hbox{\lower4pt\hbox{$\sim$}}}\hbox{$>$}}}}

\begin{document}
\thesaurus {08(08.16.5; 08.12.1; 08.12.2);10(10.15.2 Pleiades)}
\title {Brown Dwarfs in the Pleiades Cluster. II. $J$, $H$ and $K$ photometry
\thanks{Based on observations made with the William Herschel Telescope, 
operated on the island of La Palma by the Royal Greenwich Observatory in 
the Spanish Observatorio del Roque de los Muchachos of the Instituto de 
Astrof\'\i sica de Canarias (IAC), Spain.}
}

\author {M.R. Zapatero Osorio 
      \and E.L. Mart\'\i n
      \and R. Rebolo}
\offprints {M.R. Zapatero Osorio, e-mail: mosorio@iac.es}
\institute {Instituto de Astrof\'\i sica de Canarias,
            V\'\i a L\'actea s/n, E-38200 La Laguna, Tenerife, Spain}
\date {Received date; accepted date}

\maketitle

\begin{abstract}
We have obtained near-infrared observations of some of the faintest objects 
so far known towards the Pleiades young stellar cluster, with the purpose of 
investigating the sequence that connects cluster very low-mass stars with 
substellar objects. We find that infrared data combined with optical 
magnitudes are a useful tool to discriminate cluster members from foreground 
and background late-type field stars  contaminating optical surveys. 
The bottom of the Pleiades sequence is clearly defined by the faint 
HHJ~objects as the very low-mass stars approaching the substellar limit, 
by the transition object PPl~15, which will barely ignite its hydrogen 
content, and by the two brown dwarfs Calar~3 and Teide~1. 

Binarity amongst cluster members could account for the large dispersion 
observed in the faint end of the infrared colour-magnitude diagrams.
 Two objects 
in our sample, namely HHJ~6 and PPl~15, are overluminous compared to other 
members, suggesting a probable binary nature. We have reproduced the 
photometric measurements of both of them by combining the magnitudes of 
cluster very low-mass stars and brown dwarfs and using the most recent 
theoretical evolutionary tracks. The likely masses of the components are 
slightly above the substellar limit for HHJ~6, while they are 0.080 and 
0.045$\pm$0.010~$M_{\odot}$ for PPl~15. These masses are consistent with 
the constraints imposed by the published lithium observations of these 
Pleiads. 

We find a single object infrared sequence in the Pleiades cluster connecting
very low-mass stars and brown dwarfs. We propose that the substellar mass 
limit ($\sim$0.075~$M_{\odot}$) in the Pleiades ($\sim$120~Myr) takes place 
at absolute magnitudes $M_{\rm I}$~=~12.4, $M_{\rm J}$~=~10.1, 
$M_{\rm H}$~=~9.4 and $M_{\rm K}$~=~9.0 (spectral type M7). Cluster members 
fainter by 0.2~mag in the $I$-band and 0.1~mag in the $K$-band should be 
proper brown dwarfs. The star-brown dwarf frontier in the Hyades 
cluster (600~Myr) would be located at $M_{\rm I}$~=~15.0, $M_{\rm J}$~=~11.6, 
$M_{\rm H}$~=~10.8 and $M_{\rm K}$~=~10.4 (spectral type around M9). Fon an 
age older than 1000~Myr we estimate that brown dwarfs are fainter than 
$M_{\rm K}$~=~10.9 (spectral type later than M9.5).

\keywords {Stars: pre-main sequence -- Stars: late-type -- Stars: 
low-mass, brown-dwarfs -- Open clusters: Pleiades}

\end{abstract}

\section {Introduction}
The search for free-floating brown dwarfs (hereafter referred to as BDs) 
has finally been proved succesful. After years of continuous efforts to 
detect intrinsically fainter and less massive objects both in the field 
and in open star clusters (see e.g. the reviews by Jameson \cite{jameson95} 
and Reid \cite{reid95}), the recent photometric survey of a small sky area 
in the Pleiades (age 70--120~Myr) by Zapatero Osorio, Rebolo \& Mart\'\i n 
(\cite{osorio96a}) (hereafter Paper~I) has provided objects that according 
to all available evolutionary models are beyond the substellar limit (Rebolo, 
Zapatero Osorio \& Mart\'\i n \cite{rebolo95}; Mart\'\i n, Rebolo \& Zapatero 
Osorio \cite{martin96a}). The genuine BD nature of two of them has been 
confirmed by the detection of high lithium abundances in their atmospheres 
(Rebolo et al. \cite{rebolo96}). The re-appearance of lithium in objects at 
the bottom of the Pleiades main sequence (MS) has helped to empirically 
define the borderline that separates very low-mass stars from BDs. At present, 
both sides of the substellar limit are populated with several objects which 
may allow us to establish qualitative and quantitative differences between 
them and also with respect to other field objects with similar spectral 
types. Mart\'\i n et al. (\cite{martin96a}) have started to approach this 
issue using intermediate-resolution optical spectroscopy. In this paper, we 
examine this problem using near-infrared photometry. We have obtained $J$, $H$ 
and $K$ magnitudes for some Pleiades very low-mass stars selected from the 
literature, as well as for the most interesting objects in the optical survey 
presented in Paper I. We will focus here on the utility of combining both 
optical and infrared photometry to identify true Pleiades BDs, disentangle 
binarity effects and discriminate cluster members from contaminants that 
appear in optical surveys. These observations are mainly aimed at  
characterizing the cluster substellar borderline and studying the infrared 
sequence that connects Pleiades least massive stars with BDs.

\section {Sample selection. Observations}
Our sample is presented in Table~1 where we have avoided the term ``Pleiades'' 
for those objects taken from Paper~I. All objects in the sample have measured 
spectral types equal or later than M6, and they could be located around the 
substellar limit in the Pleiades cluster according to their $I$ magnitudes and 
($R-I$) colours. PPl~15 (discovered by Stauffer, Hamilton \& Probst 
\cite{stauffer94}) is the M6.5-type Pleiad that at present defines the 
borderline which separates very low-mass stars from BDs in the cluster. 
This is based on the fact that PPl~15 has preserved some of its initial 
lithium content. The comparison with theoretical models (Basri, Marcy \& 
Graham \cite{basri96}; Rebolo et al. \cite{rebolo96}) yields that the most 
likely mass of PPl~15 is $\sim$0.08~$M_{\odot}$. HHJ~2 (M6.5), HHJ~3 (M6) 
and HHJ~6 (M6.5) are proper motion members taken from Hambly, Hawkins \& 
Jameson (\cite{hambly93}). Their optical-infrared magnitudes and spectral 
types (Steele \& Jameson \cite{steele95}) place these objects close to the 
substellar limit, but still on the stellar domain. HHJ~3 has depleted its 
lithium (Marcy, Basri \& Graham \cite{marcy94}). In fact, none of the HHJ 
objects in our sample shows lithium in their atmospheres (Basri 1996, private 
communication), implying that all of them have masses larger than 
0.08~$M_{\odot}$ and therefore, they are among the least massive stars in the 
cluster. There are no lithium observations available for PPl~1 (M6.5) 
(Stauffer et al. \cite{stauffer89}), but it has a photometry very similar to 
that of HHJ~3, and a radial velocity consistent with membership. The 
remaining objects in our sample are taken from the optical survey of 
Paper~I from which two genuine BDs have emerged, namely Teide~1 (M8) and 
Calar~3 (M8). Both have successfully passed the lithium test (Rebolo et al. 
\cite{rebolo96}). According to their radial velocities and spectral types, 
Calar~1 (M9), Calar~2 (M6), Calar~5 (M6.5) and Roque~1 (M7) seem to be 
non-members of the Pleiades (Mart\'\i n et al. \cite{martin96a}). The 
infrared photometry will shed new light on their membership in the cluster.

The $J$, $H$ and $K$ observations were performed in 1996 February~9 at the 
4.2~m William Herschel Telescope (WHT, Observatorio del Roque de los Muchachos 
on the Island of La Palma, Spain), using the WHIRCAM infrared camera 
equipped with an InSb (256$\times$256) detector array. This 
detector provided 0''.24 pixels and a field of view of 
$\sim$1$'$$\times$1$'$. The standard $J$, $H$ and $K'$ infrared filter 
set was used. The total integration time per object and per filter was 
50~s, where each final image actually consisted of 5 co-added 10~s 
exposures. Raw data were processed using standard techniques within 
the IRAF\footnote{IRAF is distributed by National Optical Astronomy 
Observatories, which is operated by the Association of Universities 
for Research in Astronomy, Inc., under contract with the National 
Science Foundation.} environment. Dark frames were first substracted 
from the images before combining them in order to obtain good 
flat-fields for each filter. Then, individual frames were divided by 
the normalized, mean flat-field.  

The photometric analysis was carried out using routines
within DAOPHOT. Weather conditions during the campaign were always
fairly photometric. Instrumental aperture magnitudes were corrected
for atmospheric extinction and transformed into the UKIRT system using
observations at different airmasses of the standards FS~9 and FS~12
(Casali \& Hawarden \cite{casali92}). The $K'$ data were converted to
the $K$-band magnitudes using the transformation equation
given in Wainscoat \& Cowie (\cite{wainscoat92}). We estimate that the
accuracy in the final magnitudes is 0.05~mag at each wavelength.
The final $J$, $H$ and $K$ photometry of the sample is provided in Table~1. 
Our magnitudes are in good agreement within $\pm$0.1~mag with those found 
in the literature for the objects in common (HHJ and PPl objects), except 
for the $K$-band of PPl~15, where a difference of 0.18~mag is observed with 
respect to Basri et al.'s (\cite{basri96}) data in the sense that our result 
is brighter. We have only a $J$ measurement for HHJ~3; for completeness in 
Table~1, $H$ and $K$ magnitudes in the UKIRT system were taken from Steele, 
Jameson \& Hambly (\cite{steele93}). 

\begin{table}
\caption[]{Infrared photometry}
\begin{center}
\begin{tabular}{lcccccc}
\hline
Name    & SpT$^a$&  $J$  &  $H$  &  $K$  & $I^{b}-K$ & $R^{c}-I$\\
\hline		
HHJ~6   & M6.5   & 14.77 & 14.08 & 13.59 & 3.41      & 2.35\\
HHJ~2   & M6.5   & 15.33 & 14.71 & 14.31 & 2.99      & 2.08\\
HHJ~3$^{d}$&M6   & 15.21 & 14.50 & 14.23 & 3.18      & 2.15\\
PPl~1   & M6.5   & 15.44 & 14.68 & 14.32 & 3.21      &     \\
PPl~15  & M6.5   & 15.36 & 14.66 & 14.14 & 3.66      & 2.25\\
Calar~1 & M9     & 16.01 & 15.41 & 14.95 & 3.23      & 2.93\\
Roque~1 & M7     & 16.22 & 15.65 & 15.21 & 3.22      & 2.50\\
Calar~2 & M6     & 16.55 & 15.81 & 15.55 & 3.11      & 2.50\\
Calar~3 & M8     & 16.29 & 15.45 & 14.94 & 3.79      & 2.54\\
Teide~1 & M8     & 16.37 & 15.55 & 15.11 & 3.69      & 2.74\\
Calar~5 & M6.5   & 17.07 & 16.23 & 15.88 & 3.13      & 2.33\\
\hline
\end{tabular}
\end{center}
NOTES. \\
$^{a}$ Spectral types for the HHJ objects are taken from Steele 
\& Jameson (\cite{steele95}), whereas for the remaining objects 
spectral types come from Mart\'\i n et al. (\cite{martin96a}). \\
$^{b}$ $I$ magnitudes come from Steele \& Jameson (\cite{steele95}) 
for the HHJ~objects; from Stauffer et al. (\cite{stauffer89}) for PPl~1; 
from Stauffer et al. (\cite{stauffer94}) for PPl~15; from Paper~I for 
Roque~1, Teide~1 and Calar~objects. \\
$^{c}$ $R$ magnitudes are taken from Hambly et al. (\cite{hambly93}) 
for the HHJ objects; from Paper~I for PPl~15, Roque~1, Teide~1 and 
Calar objects.\\  
$^{d}$ $H$ and $K$ magnitudes for HHJ~3 are taken from Steele et
al. (\cite{steele93}). 
\end{table}

\section{Infrared diagrams and cluster membership}
The $K$ vs ($I-K$) colour-magnitude diagram for our sample (filled symbols) 
and some of the Pleiades least massive objects discovered to date (open 
symbols) is shown in Fig.~1. We also plot previously known proper motion 
cluster members observed at these wavelengths by Steele et al. 
(\cite{steele93}) using the revised $I$ photometry published in Steele \& 
Jameson (\cite{steele95}). We have also included the BD candidates proposed 
by Stauffer et al. (\cite{stauffer89}) and Williams et al. 
(\cite{williams96}). Different symbols are used for clarity and no 
de-reddening has been applied. The solid line is an 
averaged MS derived from the photometry of field stars 
taken from Leggett (\cite{leggett92}), shifted to an assumed 
Pleiades distance modulus of ($m-M$)$_{\circ}$~=~5.53~mag and adopting 
$A_{\rm V}$~=~0.12~mag (Crawford \& Perry \cite{crawford76}); 
$A_{\rm I}$ and $A_{\rm K}$ were then derived using the relationships 
of Rieke \& Lefobski (\cite{rieke85}).

\begin{figure}
\psfig{figure=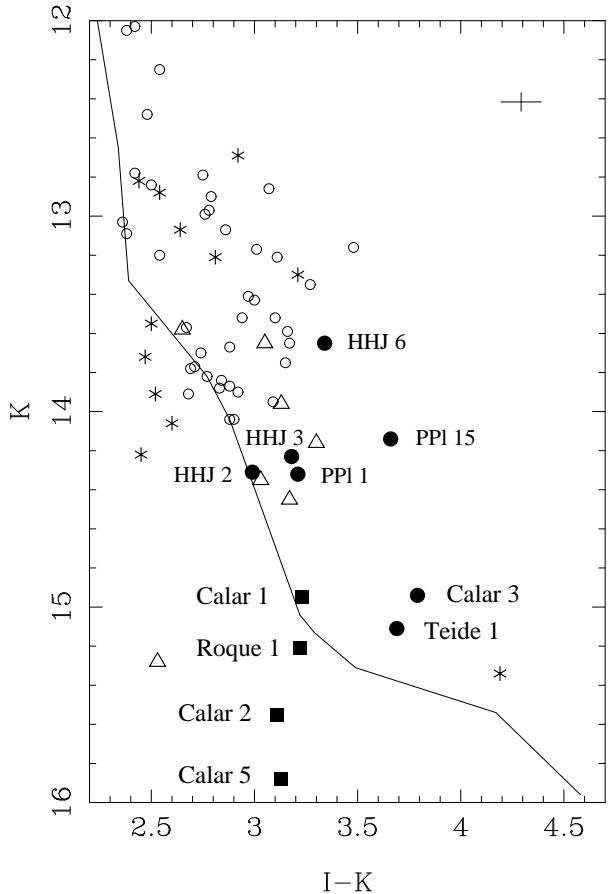,width=8cm,angle=0.0}
\caption[]{$K$ vs ($I-K$) colour-diagram for our sample and other Pleides 
low-mass objects. Filled circles denote cluster members whereas filled 
squares denote non-members (see text for explanation). Asterisks stand for 
the BD candidates of Williams et al. (1996), and triangles for those of 
Stauffer et al. (1989). Proper motion HHJ objects (data taken from Steele 
et al. 1993, and Steele \& Jameson 1995) are shown as open circles. The 
line represents the MS as derived from the photometry of field late-type 
dwarf stars (Leggett 1992), shifted to the Pleiades distance and reddening. 
Error bars for our photometry are indicated at the top right corner}
\end{figure}

Only objects with spectral types later than M3 (($I-K$) $\ge$ 2.2) are 
displayed in Fig.~1. As expected, young Pleiades members have not reached 
the field MS (Stauffer \cite{stauffer84}; Steele \& Jameson \cite{steele95}), 
thus lying above it. This is basically what is observed except for a few 
exceptions, like PPl~3 (Stauffer et al. \cite{stauffer89}), also named as 
JS1 (Jameson \& Skillen \cite{jameson89}),  which is located 
1.7~mag below the MS line. The infrared photometry rules out its membership 
in the cluster, a fact that is consistent with our proper motion measurement 
in Paper~I. Five very low-mass stars proposed by Williams et al. 
(\cite{williams96}) around ($I-K$)~=~2.5 and $K\sim14$ lie also below the 
field MS and are indeed fainter in the $K$-band than proper motion Pleiades 
of the same colour, suggesting that they are unlikely members based on their 
infrared photometry. 

Two other objects in our sample, Calar~2 and Calar~5, are clearly located 
below the MS by $\sim$1~mag in Fig.~1. Both of them have infrared colours that 
resemble those of field stars of similar spectral types within the errors in 
photometry and classification (e.g., see the ($I-K$) vs spectral type diagram 
of Fig.~2). Although found with ($R-I$) values and radial velocities 
consistent with other cluster members, Calar~2 and Calar~5 are non-members 
on the basis of their infrared photometry, confirming the suspicions 
of Mart\'\i n et al. (\cite{martin96a}). These objects do not fit in the 
sequence described by the cluster members in Fig.~4 in Mart\'\i n et al.'s
work, being fainter in $I$ than expected for Pleiads of these spectral types. 
Calar~2 and Calar~5 are therefore likely background reddened low-mass stars. 
Infrared observations allow an easy discrimination of contaminants that could 
arise in optical photometric surveys from true Pleiads.

Roque~1 appears very close to the MS in Fig.~1, but slightly 
below it implying that it is a likely non-member. By inspecting Fig.~2 
and allowing for the photometric uncertainties, 
this object fits in the ($I-K$) colour-spectral type diagram for 
solar metallicity field stars. From the comparison of its photometry and 
spectral type with the absolute magnitudes published in Kirkpatrick \& 
McCarthy (\cite{kirkpatrick94}) we conclude that this object may be located 
at approximately the Pleiades distance. Nevertheless, given its high radial 
velocity ($v_{\rm r}$~=~151$\pm$30~km s$^{-1}$), it was suggested in 
Mart\'\i n et al. (\cite{martin96a}) that 
Roque~1 could be part of the old galactic population, but neither 
its $R$, $I$ and $J$, $H$, $K$ colours nor its optical spectrum show any 
evidence of significant low-metallicity. Roque~1 still awaits further 
investigation that confirms its membership in an old population.

Calar~1 sits on the MS in Fig.~1. However, its location is highly unexpected 
given its spectral type and ($R-I$) colour. This object was found to be the 
reddest BD candidate in the optical survey of Paper~I with ($R-I$)~=~2.9, 
and surprisingly, it appears rather blue in the infrared (see Fig.~2). 
Apart from being one of the few M9 dwarfs known in the whole sky, 
it is the first one to our knowledge that behaves so rarely at infrared 
wavelengths. Leggett (\cite{leggett92}) states that for very cool dwarfs no 
obvious metallicity effects are discernable in optical diagrams, but that 
they become apparent in infrared plots. Metal-poor objects are well known 
to have bluer colours than solar metallicity ones. Calar~1 may be 
a metal-deficient dwarf located foreground to the Pleiades, a fact that 
is supported by its high radial velocity
($v_{\rm r}$~=~85$\pm$30~km s$^{-1}$, Mart\'\i n et al. \cite{martin96a}). 
The proper motion and parallax determinations would help to understand the 
nature of this object. Calar~1, Calar~2, Calar~5 and Roque~1 are excluded 
from further discussion in this paper as very unlikely members of the cluster.

\begin{figure}
\psfig{figure=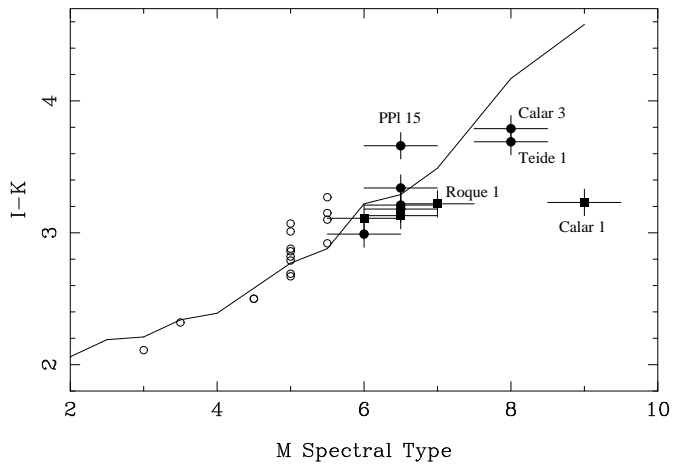,width=8.8cm,angle=-90}
\caption[]{($I-K$) vs spectral type diagram for our sample. Symbols are as 
in Fig.~1. Spectral types have been taken from Steele \& Jameson (1995), 
Mart\'\i n et al. (1994), and Mart\'\i n et al. (1996). The line represents 
the relationship for solar metallicity field stars (Kirkpatrick \& McCarthy 
1994). Error bars are half a spectral subclass (as stated by the authors), 
and $\pm$0.1~mag for the photometric colour. Some objects have been labeled 
for clarity}
\end{figure}

The remaining objects in our sample (HHJ~2, HHJ~3, HHJ~6, PPl~1, PPl~15, 
Teide~1 and Calar~3) do follow the sequence described by the proper motion 
cluster members in Fig.~1. Particularly, Teide~1 and Calar~3 define the 
present faint end of the Pleiades, which lies  0.3--0.45~mag above the MS 
in the $K$-band. Objects with similar photometry should be considered 
as probable BDs in the cluster. The object in Williams et al. 
(\cite{williams96}) labelled as No.~13 could be a BD based on its photometry.
However, we have recently obtained deep images in the $I$-band which show  
that it is likely an extense object.

\subsection{Broad-band energy distributions}
We have combined  optical ($R$ and $I$) and infrared magnitudes of the 
Pleiades objects in our sample in order to obtain the general shape of their 
energy distributions and compare them with those of field stars with 
similar spectral types. We adopted the absolute magnitudes for M 
dwarfs given in Kirkpatrick \& McCarthy (\cite{kirkpatrick94}) as 
representative of the field, and the distance modulus and extinction 
of the Pleiades to convert the observed magnitudes to absolute 
magnitudes for the cluster members. The zero-magnitude fluxes and 
center-wavelengths (Cousins for $R$ and $I$, UKIRT for $J$, $H$ and $K$) 
are taken from Mead et al. (\cite{mead90}).

In Fig.~3 we plot the final broad-band spectral distributions for PPl~1, 
PPl~15, Teide~1, and Calar~3. PPl~1 is the only object in our sample which 
has no $R$ magnitude available. We have estimated it using the mean ($R-I$) 
colour of the other three M6.5 Pleiads in our sample (the mean coincides 
with that given by Kirkpatrick \& McCarthy \cite{kirkpatrick94} for a field 
star of the same spectral type). Overplotted are the distributions of stars 
from M5.5 to M9 (for spectral types M8 and M9 Kirkpatrick \& McCarthy 
\cite{kirkpatrick94} do not provide data at the $R$-band). This plot clearly 
shows how important infrared emission is to study these extremely late-type 
objects since most of their bolometric flux is emitted at these wavelengths. 
Less than a few per cent is emitted blueward of the $R$-band. 

Given a spectral type, a Pleiades member having the same effective 
temperature and hence a very similar energy distribution than a field 
star, should emit more at each wavelength due to its youth; i.e., it 
should move up along the flux-axis in Fig.~3. PPl~1 and PPl~15 appear 
to have energy distributions that resemble within uncertainties in flux 
conversion and observed magnitudes those of the old M6--M7 stars, 
except that the formers (which have similar optical spectral types) 
are twice (0.3~dex) more luminous. Teide~1 and Calar~3 show cooler 
energy distributions that are much alike to M8-type field stars, but 
the two Pleiades BDs are overluminous by a factor 1.6 (0.2~dex). 
These results agree with expectations for cluster members. 

\begin{figure}
\psfig{figure=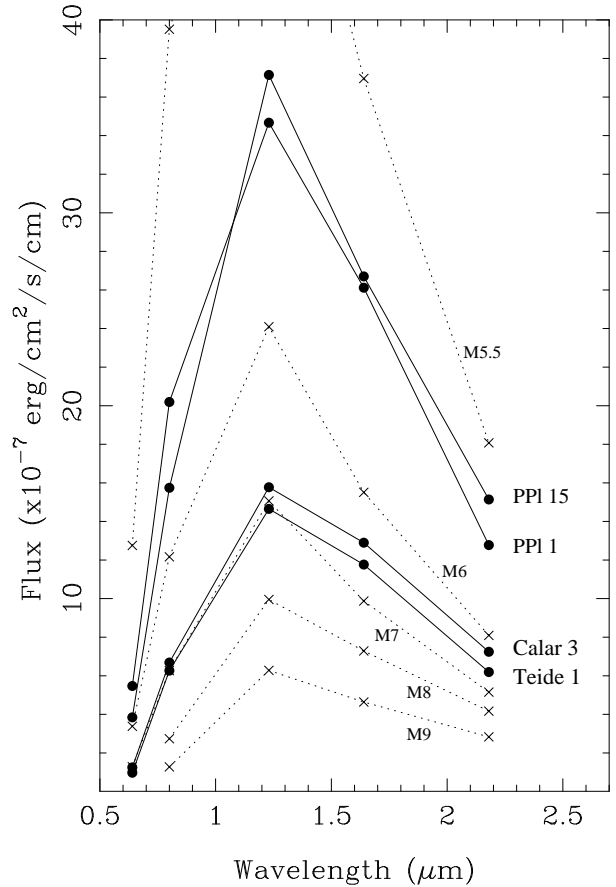,width=8cm,angle=0.0}
\caption[]{Broad-band energy distributions at $R$, $I$, $J$, $H$ and $K$ 
wavelengths. Pleiades members are denoted as circles and full lines. 
The spectral distributions of M5.5--M9 field dwarfs are also plotted for 
comparison (crosses and dotted lines). Errors in flux conversion could be 
as large as 20\%.}
\end{figure}

The accurate determination of  bolometric 
luminosities for our objects would require  
the use of spectroscopic data covering the whole optical and infrared regions. 
Integrating the broad-band flux distributions always overestimates the 
luminosity, because the spectra of these late-type objects are dominated 
by deep molecular and absorption features that are not precisely described 
by the photometry. The overestimation (less than 10\% \ according to 
Reid \& Gilmore \cite{reid84} and Tinney et al. (\cite{tinney93}) 
seems to be constant through all the spectral types, although perhaps slightly 
larger for M9 stars. This allow us to be confident on the fact that relative 
luminosities within our sample are not strongly influenced by molecular 
absorption effects. We have integrated the flux distributions of our Pleiades 
members and evaluated the differences in luminosity with respect to Teide~1 
(the least luminous object in our sample). The resulting values are provided 
in Table~2. We estimate that they are affected by an error of $\pm$0.04~dex 
which mainly comes from uncertainties in the photometry. 

\begin{table}
\caption[]{Differences in log~($L/L_{\odot}$) relative to Teide~1}
\begin{center}
\begin{tabular}{cccccc}
\hline
HHJ~6 & HHJ~2 & HHJ~3 & PPl~1 & PPl~15 & Calar~3 \\
\hline
0.63  & 0.41  & 0.45  & 0.36  & 0.37   & 0.05    \\
\hline
\end{tabular}
\end{center}
\end{table}

In the Pleiades cluster stars of spectral type M6.5 are 
2.29$\pm$0.25~times (0.36$\pm$0.04~dex) more luminous than M8-type objects. 
We have computed the mean luminosities for field stars of the same spectral 
types averaging the data from Tinney et al. (\cite{tinney93}) and 
Bessell \& Stringfellow (\cite{bessell93}), and found that M6.5 field stars 
are overluminous by a factor 1.78 (0.25~dex) with respect to M8 stars. It is 
remarkable that the factor for the Pleiades is larger by $\sim$0.1~dex. This 
offset in luminosity takes into account differences in radii. The fact that 
the ratio $R_{\rm M6.5}/R_{\rm M8}$ is greater in the Pleiades cluster than 
in the field could be explained by a larger mass difference between a M6.5 
object and a M8 object in the former than in the latter. This result was 
expected from theoretical evolutionary models. It would be desirable to 
define a consistent spectral type classification using objects that belong 
to a cluster to take advantage of the coeval formation as well as of the 
homogeneous metallicity.

\section{Binarity amongst the faintest Pleiades members}
One of the most noteworthy features in the colour-magnitude diagram of Fig.~1 
is the large scatter observed among the Pleiades objects. As a possible 
explanation it was suggested that an age spread could account for such an 
effect (Eggen \& Iben \cite{eggen89}; Steele et al. \cite{steele93}). However, 
this is a controversial subject (e.g. Stauffer, Liebert \& Giampapa 
\cite{stauffer95}). 
Recently, Steele \& Jameson (\cite{steele95}) have argued that 
the scatter could be interpreted in terms of a single and a binary 
star sequence, both of the same age. These authors found that 46\% \ 
of the fainter HHJ stars are likely binary systems, a figure which 
compares well with those given for low-mass stars (Bessell \& Stringfellow 
\cite{bessell93}, and references therein). According to Steele \& 
Jameson (\cite{steele95}), the binary sequence of M stars at the age 
of the Pleiades is expected to lie $\sim$0.75~mag above the single 
sequence. It is obvious from Fig.~1 that two Pleiads (HHJ~6 and PPl~15) 
in our sample are too bright by a similar quantity in the $K$-band, 
thus suggesting that they are binary sistems. We will take advantage of 
what is already known for these two cluster members and try to estimate 
the most likely masses of the components. We will show that it is possible 
to reproduce the observed magnitudes and colours of HHJ~6 and PPl~15 by 
co-adding those of other members of different masses located on the sequence 
of likely single objects, which is defined by HHJ~2, HHJ~3, PPl~1, Calar~3 
and Teide~1. These members can be used as photometric and mass templates.

HHJ~3 does not show lithium in its atmosphere implying that its mass is 
greater than 0.08~$M_{\odot}$. PPl~1 has a very similar photometry both 
in the optical and in the infrared, and therefore it could be considered 
as a ``twin'' of HHJ~3. For Calar~3 and Teide~1, Rebolo et al. 
(\cite{rebolo96}) inferred a likely mass of 0.055$\pm$0.015~$M_{\odot}$ 
based on the comparison of their lithium abundance and luminosity with the 
theoretical models of Chabrier, Baraffe \& Plez (\cite{chabrier96}) that 
predict the destruction of this element as a function of time. The authors 
adopted a cluster age of 120~Myr (see Basri et al. \cite{basri96} for a 
further discussion). For the following analysis we have taken the same age 
as well as the theoretical evolutionary tracks of Chabrier et al. 
(\cite{chabrier96}) (those based on the 'NextGen' model atmospheres of 
Allard \& Hauschildt \cite{allard96}). We have computed the magnitude 
offsets for different masses in the range 0.10--0.04~$M_{\odot}$ relative 
to an object of 0.055~$M_{\odot}$, and applied them to the photometry of 
Teide~1. We have built a grid of apparent magnitudes that describe very 
low-mass single stars and BDs in the cluster. The comparison of the 
photometry of HHJ~3 and PPl~1 with the grid yields a mass of 
0.09$\pm$0.01~$M_{\odot}$, consistent with the high destruction of lithium 
observed in them.

HHJ~6 is massive enough to be a star since it has depleted its lithium 
content. Its magnitudes are brighter than those of HHJ~3 
and PPl~1 which have the same spectral type, whereas their colours are not 
different within error bars (see Fig.~2). This suggests that if HHJ~6 
is a binary system, the components should have roughly equal masses. 
Steele \& Jameson (\cite{steele95}) and Steele et al. (\cite{steelej95}) 
obtained the optical and infrared spectra, respectively, of this Pleiad 
and claimed that their spectroscopic data do not show any evidence for a 
companion of different mass. Following the same approach described above and 
taking the 0.09~$M_{\odot}$ stars as templates, we find that the combination 
of two objects of 0.085$\pm$0.010~$M_{\odot}$ each reproduces well the 
magnitudes and colours measured for HHJ~6. The number of possible 
pairs that comply with the purpose of obtaining the photometry of a 
presumed binary object is not unique, but the current spectroscopic 
knowledge on the candidate constrains the result. Nevertheless, given the 
error bars, other combinations of objects with masses around 0.09 and 
0.08~$M_{\odot}$ are possible (e.g., 0.090 and 0.085~$M_{\odot}$; 
0.09 and 0.08~$M_{\odot}$). 

\subsection{A likely brown dwarf companion to PPl~15}
PPl~15 is a M6.5 Pleiad that shows at infrared wavelengths too red colours 
for its spectral type (see Fig.~2), even taking into account the 
difference in the $K$-band magnitude between Basri et al.'s (\cite{basri96}) 
measurement and ours. This object, however, fits well the single sequence of 
cluster members in the $V$ versus ($V-I$) (Stauffer et al. \cite{stauffer94}) 
and in the $I$ versus ($R-I$) (Paper~I) diagrams. A possibility that could 
explain the observed excess in the infrared colours of PPl~15 is the existence 
of a residual extinction dust shell around this object. According to previous 
$J$, $H$ and $K$ data (Stauffer \cite{stauffer82}, \cite{stauffer84}; Steele 
\& Jameson \cite{steele95}), low-mass Pleiades stars have colours that 
resemble those of field dwarfs. This suggests that if a circumstellar disc 
exists its size should be small and hence, it hardly contributes to the total 
luminosity of the object. The anomalous colours of PPl~15 may be attributable 
to binarity, however. The supposed secondary must be significantly 
less massive than the primary. Its contribution to the total integrated flux 
would be hence more important at redder wavelengths. The flux emitted at $J$, 
$H$ and $K$ by PPl~15 is larger than that emitted by PPl~1, occurring just 
the opposite at $R$ and $I$ wavelengths as shown in Fig.~3. Following the 
technique of the relative differences in colour previously described, we 
find that a combination of objects of 0.080 and 0.045$\pm$0.010~$M_{\odot}$ 
matches the photometry of PPl~15. 

It is of interest to calculate the contribution of such a low-mass companion 
to the total luminosity of the PPl~15 system as well as to the observed 
magnitude in each filter. According to Rebolo et al. (\cite{rebolo96}), 
the luminosity of PPl~15 is 
log~$L/L_{\odot}$~=~--2.80$\pm$0.10, about 20\% \ of it could come from the 
secondary. Its contribution to the $R$-band is only 5\%. Thus, the effect of 
this companion on the ``unresolved'' optical spectrum obtained by Basri et 
al. (\cite{basri96}) is nearly undetectable, in particular implying that the 
observed lithium line at $\lambda$670.8~nm is due to the primary component, 
for which we have estimated a mass just on the substellar limit. To date, 
there is no report on the variabililty of the radial velocity of PPl~15. It 
would be desirable to obtain near infrared spectroscopy of this object in 
order to test its possible binarity. At these wavelengths (1--2.5~$\mu$m) the 
low-mass companion is expected to contribute with 25--30\% \ of the total 
flux, and hence its effects on molecular bands such as CO and H$_{2}$O should 
be noticeable (Steele et al. \cite{steelej95}). 

\section{The infrared sequence from very low-mass stars to brown dwarfs}
The final $K$ versus ($I-K$) diagram for the least massive Pleiades members 
known so far is displayed in Fig.~4. We have only plotted objects for which 
cluster membership has been established by other means than photometry, such 
as proper motion and radial velocitiy measurements, and/or the presence of 
lithium in their spectra. Superposed on the data we have drawn theoretical 
evolutionary tracks of Baraffe et al. (\cite{baraffe95}) based  on the 'Base' 
model atmospheres of Allard \& Hauschildt (\cite{allard95}), and those of 
Chabrier et al. (\cite{chabrier96}) based on the most recent 'NextGen' model 
atmospheres of Allard \& Hauschildt (\cite{allard96}) for solar metallicity 
and ages of 70 and 120~Myr. We have also included the isochrones for 70~Myr 
of Burrows et al. (\cite{burrows93}) (Model~X) and D'Antona \& Mazzitelli 
(\cite{dantona94}) (Table~5). Baraffe et al. and Chabrier et al. have 
estimated magnitudes ($V$, $R$, $I$ in the Cousins system and $J$, $H$, $K$ 
in the CIT system) as a function of mass and age. We have used the equations 
given in Leggett (\cite{leggett92}) to convert the infrared magnitudes into 
the UKIRT system. Burrows et al. (\cite{burrows93}) and D'Antona \& Mazzitelli 
(\cite{dantona94}) do not provide magnitudes for their models; thus, we have 
computed them using the ($I-K$) and $K$ bolometric correction versus effective 
temperature relationships that can be obtained from Tables~2, 6 and~7 of 
Leggett et al. (\cite{leggett96}). 

\begin{figure}
\psfig{figure=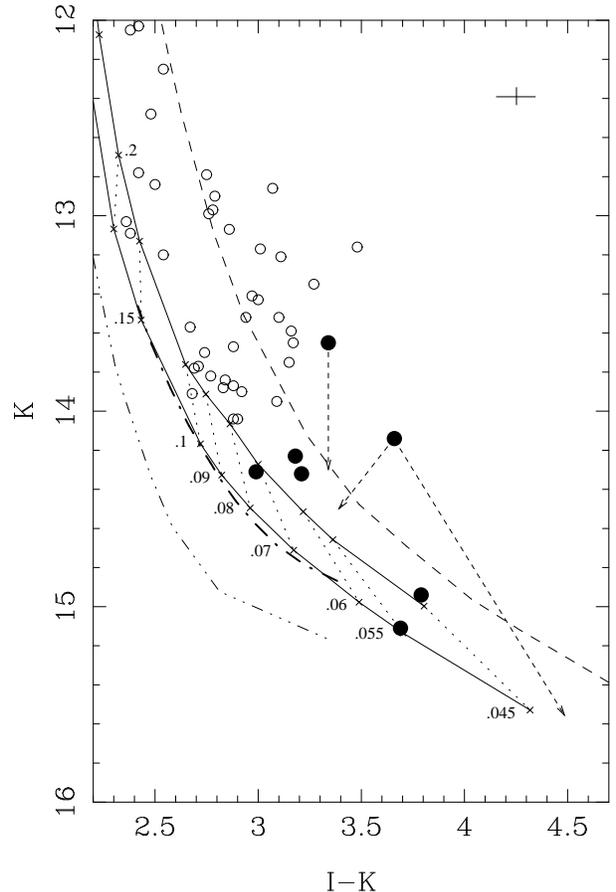,width=8cm,angle=0}
\caption[]{$K$ vs ($I-K$) colour-diagram for the Pleiades members. Symbols 
are as in Fig.~1. Superimposed to the data are the isochrones of Baraffe et 
al. (1995) based on the 'Base' models atmospheres (Allard \& Hauschildt 1995, 
dashed line, 120~Myr), and those of Chabrier et al. (1996) based on the 
``NextGen'' model atmospheres (Allard \& 
Hauschildt 1996, solid lines, 70 and 120~Myr). Numbers denote masses in solar 
units for these models, and evolutionary tracks are shown as dotted lines. 
Isochrones of Burrows et al. (1993) and D'Antona \& Mazzitelli (1994) are 
denoted by a dot-dashed line and by a dot-dot-dot-dashed line, respectively, 
both for an age of 70~Myr. The mass range depicted for these models is 
0.15--0.04~$M_{\odot}$. An indication of the location of the likely components 
of HHJ~6 and PPl~15 is given by dashed arrows}
\end{figure}

\begin{table*}
\caption[]{Absolute magnitudes defining the substellar limit (0.080--0.075~$M_{\odot}$, solar metallicity) at different ages}
\begin{center}
\begin{tabular}{ccccccc}
\hline
Age            &$M_R$ &$M_I$ &$M_J$ &$M_H$ &$M_K$ & SpT \\
(Myr)          &      &      &      &      &      &     \\
\hline
Pleiades (120) & 14.4--14.7 & 12.2--12.4 & 10.0--10.1 &  9.3--9.4 &  8.9--9.0  & M6.5--M7\\
Hyades (600)   & 17.2--17.9 & 14.5--15.0 & 11.4--11.6 & 10.6--10.8 & 10.3--10.4 & M8.5--$\qqap$M9\\
Field (1000)   & 18.0--18.9 & 15.1--15.8 & 11.8--12.1 & 10.9--11.2 & 10.6--10.9 & $\qqap$M9--$>>$M9\\
Field (5000)   & 18.9--20.5 & 15.8--17.2 & 12.1--13.0 & 11.3--12.1 & 10.9--11.8 & $>>>$M9\\
\hline
\end{tabular}
\end{center}
\end{table*}

All isochrones reproduce qualitatively the general trend of the observational 
data. However, within the age range assumed for the Pleiades (70--150~Myr), 
the models of Burrows et al. (\cite{burrows93}) and D'Antona \& Mazzitelli 
(\cite{dantona94}) predict rather blue ($I-K$) colours and/or faint $K$ 
magnitudes. Only for very young ages, those isochrones would fit the Pleiades 
data, implying that cluster members like the faintest HHJ objects should have 
preserved lithium, clearly in desagreement with the lithium observations of 
these stars. This discrepancy for very low-masses was also noted in terms 
of effective temperature and luminosity in the $\alpha$~Persei cluster by 
Zapatero Osorio et al. (\cite{osorio96b}). The 70~Myr evolutionary track of 
Burrows et al. (\cite{burrows93}) seems to coincide with that of the `NextGen' 
120~Myr isochrone of Chabrier et al. (\cite{chabrier96}). For a given age, 
Burrows et al.'s models predict effective temperatures that are hotter by 
150--100~K than those predicted by Chabrier et al., and luminosities that 
are slightly brighter by 0.03--0.015~dex in the mass range in common for 
both theoretical models. The differences between the two sets of isochrones 
of Baraffe et al. (`Base') and Chabrier et al. (`NextGen') are remarkable, 
e.g., given an age and a mass the ($I-K$) offset could be as large as 
0.5~mag. The `Base' model atmospheres have been improved recently by 
including more accurate molecular line lists, yielding the `NextGen' models. 
These are claimed to reproduce the observations better than previous sets of 
computations (see Allard et al. \cite{allard96}; Baraffe \& Chabrier 
\cite{baraffe96}; Chabrier et al. \cite{chabrier96}). However, the masses 
that are inferred from the comparison of our data with the `NextGen' models 
in Fig.~4 are lower by a factor $\sim$1.25 than those obtained from the 
lithium test. This is basically due to the apparent different age that is 
derived from the isochrones in Fig.~4 (70~Myr, the canonical age of the 
cluster), and from the combination of lithium and luminosity of objects 
close to the substellar limit (120~Myr, Basri et al. \cite{basri96}). 
Since the input molecular line lists for computing synthetic spectra for 
cool temperatures are incomplete, colours and magnitudes measured on these 
spectra are hampered by uncertainties of $\sim$0.5~mag (Chabrier et al. 
\cite{chabrier96}). On the other hand, masses obtained from lithium are 
more reliable because they are based on the simpler physics of the interior 
of fully-convective objects (Magazz\`u, Mart\'\i n \& Rebolo 
\cite{magazzu93}; Bildsten et al. \cite{bildsten96}). In order to bring 
the masses obtained from the $K$ vs $I-K$ diagram into agreement with 
those inferred from the lithium test for an age of 120~Myr, it would be 
required to shift the `NextGen' isochrone by about 0.4~mag towards redder 
colour, which is within the theoretical uncertainties estimated by Chabrier 
et al. \cite{chabrier96}.    

The sequence of very low-mass stars and BDs in the Pleiades is depicted in 
Fig.~4. Our suggested decomposition for HHJ~6 and PPl~15 is indicated. The 
location of the likely companions nicely fits the sequence of single members. 
We consider that the primary of PPl~15 defines the transition region between 
stars and BDs, i.e. masses within 0.08--0.075~$M_{\odot}$, and we propose 
that its photometry could be adopted as indicative of the location of the 
substellar limit in the Pleiades. The derived absolute magnitudes are listed 
in Table~3, where the interval of values corresponds to the above mass range. 
We have estimated the star-BD frontier for older clusters like the Hyades, 
as well as for objects of different ages in the field using the isochrones 
of Chabrier et al. (\cite{chabrier96}) ('NextGen' model atmospheres) 
normalized to the primary of PPl~15. We consider the theoretical magnitude 
differences associated to different ages in order to evolve from the Pleiades 
to 600, 1000 and 5000~Myr. We have also inferred the spectral types of the 
objects at the substellar limit --~shown in the last column of Table~3~-- 
by comparing our $I-J$, $I-K$ and $J-K$ colours with those of Kirkpatrick 
\& McCarthy (\cite{kirkpatrick94}) for field objects. We obtain fainter 
magnitudes and cooler spectral types than suggested by the mass-luminosity 
and mass-spectral type relationships proposed by Henry \& McCarthy 
(\cite{henry93}) and Kirkpatrick \& McCarthy (\cite{kirkpatrick94}) 
respectively. Our results are in better agreement with those of Baraffe 
\& Chabrier (\cite{baraffe96}). At the age of the Hyades, the substellar 
limit has moved to M9 spectral type; for example, cluster objects with the 
mass of the primary of PPl~15 might show a spectrum similar to a field M9 
dwarf, whereas those like Teide~1 would have become much cooler. Field very 
late-M dwarfs (M7--M10) are likely older than 1000~Myr and, therefore, 
most of them should be very low-mass stars rather than substellar objects. 
It is possible that photometric surveys reveal nearby young BDs with 
spectral types earlier than M9--M10 in the solar neighbourhood. Because 
of their recent formation they will 
be less numerous than stars of similar spectral types in the field, but their
relatively higher luminosity will favour detection in all-sky near-infrared 
surveys like DENIS and 2MASS. The probability of finding these field young 
BDs will depend on the mass function for very low-masses and on the mixture 
of ages (see Mart\'\i n, Zapatero-Osorio \& Rebolo \cite{martin96b}). 
The lithium test will be useful in order to establish their substellar nature.

\section{Conclusions}
We have presented near-infrared $J$, $H$ and $K$ photometry for some of the 
least luminous objects so far known in the Pleiades cluster. They are very 
low-mass stars approaching the substellar limit, transition objects, and 
massive BDs (HHJ and PPl objects, and Calar~3 and Teide~1).  In addition, 
we have also included in our sample four of the BD candidates proposed in 
Paper~I (Calar~1, 2, 4, 5 and Roque~1) in order to asses their membership 
on the basis of the infrared data. We have made use of all the available 
spectroscopic and photometric knowledge on our final list of objects to 
characterize the sequence that cluster members around the substellar 
limit and beyond do follow in the infrared colour-magnitude diagrams. 
This sequence is clearly defined by the faintest HHJ~objects, PPl~1, 
PPl~15, Calar~3 and Teide~1, lying 0.3--0.4~mag above the field MS for 
spectral types later than M6. Objects found in future surveys with similar 
photometry (optical and infrared) should be considered as likely cluster 
members. 

Undoubtely, infrared data is highly useful to discriminate those objects that 
are true Pleiades members from those that are not. Optical surveys could be 
contaminated by background objects that although fitting in the cluster 
sequence at these wavelengths, may suffer from interestellar reddening. 
As extinction affects much less infrared magnitudes, consequently these 
objects do not match the $J$, $H$ and $K$ photometry expected for the 
Pleiades (this is the case of Calar~2 and Calar~5). Foreground late-type 
stars are also possible contaminants in photometric surveys, although its 
contamination is expected to be considerably smaller. However, in the 
optical survey of Paper~I two of these field stars were detected, Roque~1 
and Calar~1, both having very high radial velocities (Mart\'\i n et al. 
\cite{martin96a}). The reddest object in optical wavelengths, Calar~1 (M9), 
appears to have rather bluer infrared colours than expected for its spectral 
type. This together with its high radial velocity suggests membership to 
an old population. 

The large dispersion observed in the infrared photometry of Pleiades members 
is likely attributable to binarity effects. Two objects in our sample, namely 
HHJ~6 and PPl~15, lie $\sim$0.7~mag above the single cluster sequence in 
infrared diagrams. Adopting an age of 120~Myr for the cluster and using the 
photometric data of very low-mass Pleiades stars and BDs as well as the 
theoretical tracks of Chabrier et al. (\cite{chabrier96}), we have reproduced 
the photometry of both, finding that the likely masses of the components are 
0.085 and 0.085$\pm$0.010~$M_{\odot}$ for HHJ~6, 0.080 and 
0.045$\pm$0.010~$M_{\odot}$ for PPl~15. These results are in fairly good 
agreement with those obtained from the lithium test. Follow-up infrared 
spectroscopic observations of PPl~15 should be made to try to detect the 
effects of the companion on the spectrum, which are expected to be detectable. 
Since the primary of PPl~15 would be exactly located at the transition region 
between stars and BDs in the Pleiades, we have used it to define the empirical 
location of the substellar limit in infrared diagrams, for which we have 
corrected the observed magnitudes from the contribution of the secondary. We 
have estimated that the star-BD frontier on the single Pleiades sequence is 
located at  magnitudes $I$~=~17.8--18.0, $J$~=~15.6--15.7, $H$~=~14.9--15.0 
and $K$~=~14.5--14.6 and at optical spectral types M6.5--M7. New Pleiades 
objects that are found fainter along the photometric sequence defined in this 
paper might be BDs indeed. Based on this result, we estimate that the 
hydrogen-burning mass limit occurs at spectral types around M9 in the Hyades 
cluster. 

\acknowledgements
{It is a pleasure to acknowledge I. Baraffe and G. Chabrier for providing 
ASCII files of their models. We thank F. Prada for his valuable support at 
the telescope. We are also indebted to M. Murphy for her careful reading of 
the manuscript and English corrections. Partial financial support was 
provided by the Spanish DGICYT project no. PB92--0434--C02.}

\end{document}